\documentclass[conference]{IEEEtran}

 \newcommand\ie{\textsl{i.e.},\ }
 \newcommand\eg{\textsl{e.g.},\ }
 
 \newcommand\etal{{\em et al.}}

\usepackage{cite}
\usepackage{underscore}
\usepackage[english]{babel}
\usepackage{blindtext}
\usepackage{lipsum}
\usepackage{algorithmic}
\usepackage{graphicx}
\usepackage{xcolor}
\usepackage{subcaption}

\usepackage{etoolbox}
 \makeatletter
 \patchcmd{\@maketitle}
   {\addvspace{0.5\baselineskip}\egroup}
   {\addvspace{-0.5\baselineskip}\egroup}
   {}
   {}
 \makeatother

\begin{document}

\title{OpenFlow-compliant Topology Management for SDN-enabled Information Centric Networks}

\author{
\IEEEauthorblockN{George Petropoulos, Konstantinos V. Katsaros, Maria-Evgenia Xezonaki}%
\IEEEauthorblockA{Intracom SA Telecom Solutions, \\
Peania, 19002, Greece\\
Email:\{geopet, konkat, maxez\}@intracom-telecom.com}
}

\maketitle

\begin{abstract}
Information-Centric Networking (ICN) has emerged as an interesting approach to overcome many of the limitations of legacy IP-based networks. However, the drastic changes to legacy infrastructure required to realise an ICN have significantly hindered its adoption by network operators. As a result, alternative deployment strategies are investigated, with Software-Defined Networking (SDN) arising as a solution compatible with legacy infrastructure, thus opening new possibilities for integrating ICN concepts in operators' networks. This paper discusses the seamless integration of these two architectural paradigms and suggests a scalable and dynamic network topology bootstrapping and management framework to deploy and operate ICN topologies over SDN-enabled operator networks. We describe the designed protocol and supporting mechanisms, as well as the minimum required implementation to realize this inter-operability. A proof-of-concept prototype has been implemented to validate the feasibility of the approach. Results show that topology bootstrapping time is not significantly affected by the topology size, substantially facilitating the intelligent management of an ICN-enabled network. 
\end{abstract}

\IEEEpeerreviewmaketitle

\section{Motivation}
Information-centric networking (ICN) and software defined networking (SDN) have emerged to address the rapid increase in traffic demands and operational requirements of business applications. However, the majority of ICN propositions suggest replacing legacy network protocols and infrastructure with information-centric alternatives. This brings severe barriers to the adoption of the ICN paradigm, as network operators resist such costly and disruptive changes. On the other hand, SDN builds on backward compatibility, facilitating adoption by network and service providers, which have already started incorporating SDN capabilities into their networks. 
The emergence of SDN creates new opportunities for deploying the novel routing mechanisms brought by the ICN paradigm, bringing its adoption by Internet stakeholders closer to reality. 

\vspace{-0.5pt}
Focusing on backward compatibility, POINT\cite{Trossen15} takes an alternative migration path that suggests the realization of legacy IP services on top of ICN and enabling the network operators to benefit from ICN core functions \eg multicast and in-network caching. The envisioned strategy includes the deployment of gateways at the edges of IP networks, to map IP-based protocols to ICN semantics, whilst supporting ICN in the network core. The network core design subsequently builds on SDN primitives to enable an IP/backwards-compatible architecture utilizing the PURSUIT~\cite{Trossen072012} publish/subscribe communication paradigm. SDN switching~\cite{Reed15} is employed for the support of a source-routing forwarding mechanism~\cite{Jokela09}, enabling stateless, fast and scalable content forwarding. 
This allows the seamless deployment of ICN topologies over SDN networks and opens a potential migration path for ICN from Internet stakeholders.

In this context, support for this transparent ICN operation heavily depends on the appropriate configuration of SDN switches, as a key enabler of the aforementioned IP-transparent flow rule matching. 
This subsequently calls for the definition of a network bootstrap and management protocol that can alleviate network operators from the cumbersome manual configuration of their SDN infrastructure, allowing for automated and scalable configuration operations. In this paper we fill this gap presenting our SDN enabled, scalable bootstrapping solution utilizing OpenFlow~\cite{OpenFlow}. We design and implement a protocol which dynamically distributes network configuration parameters to create and extend an ICN topology over an existing SDN network, without modifying existing IP or SDN protocols and services, thus achieving \textit{full backward compatibility}. 
The approach is not limited to proactive bootstrapping only, but also allows dynamic SDN switch and host attachment to an already deployed topology. An interface between the SDN and ICN control realms is introduced to support the exchange of resource allocation and monitoring messages and further realize traffic engineering and resilience functionalities. We consider the proposed solution as a key step towards the adoption of an IP compatible ICN architecture. Our solution alleviates management and configuration overheads, by enabling the automated, pro-active and re-active network topology management and bootstrapping, subsequently facilitating deployments at large scale.

\vspace{-0.5pt}

\section{Related Work}
ICN-SDN inter-operability has already attracted significant research interest. Initial work extended OpenFlow with ICN forwarding functions, as an additional feature to the provided Ethernet/IP forwarding~\cite{Suh2012}. 
NDNFlow implements a separate application layer for ICN communication traffic on top of the underlying IP and Ethernet forwarding plane, and requires an ICN-aware SDN controller module for path computation and network management~\cite{Van2015}. The CCNx daemon has to be installed onto legacy SDN switches. 
A similar strategy is followed in~\cite{Nguyen2013}, where a CCNx wrapper is implemented for OpenFlow switches. 

Other approaches base their work on the OpenFlow specification, aiming to get closer to deployment. Salsano \etal~\cite{Salsano2013} suggest overloading the IP protocol header with ICN semantics to enable SDN switches to identify ICN traffic and request the forwarding path from the SDN controller. 
Vahlenkamp \etal~\cite{Vahlenkamp2013} removes ICN awareness from the switches, decoupling ICN control and data planes. ICN packets are transferred over any IP transport protocol. Eum \etal~\cite{Eum2015} suggest a similar mechanism for federated sensor networks, embedding ICN forwarding identifiers to UDP header fields. 

Most of the proposed mechanisms either suggest overloading IP protocol semantics with content identifiers and metadata, or require the SDN controller or the switches to be involved in data plane ICN functions. Moreover, bootstrapping and management of ICN networks is not examined. Our work builds on top of the initial architectural concept proposed in~\cite{Syrivelis2012}, utilizing the stateless ICN over SDN forwarding mechanism of~\cite{Reed15} and goes beyond the state of the art by suggesting a proactive and scalable protocol for \textit{bootstrapping ICN topologies in SDN-enabled networks}. An ICN-aware SDN controller is only involved during bootstrapping of nodes, and stateless forwarding of ICN traffic is performed without disrupting legacy IP services. Furthermore, our framework enables \textit{traffic engineering and resilience management} operations for the created ICN topologies.

\section{ICN Deployment}

\subsection{Assumptions and Terminology}
In our work, the network operator is assumed to have already deployed a physical network with link-layer connectivity. The aim is to deploy and manage an ICN network over the legacy infrastructure without disrupting existing protocols and applications \ie nodes will continue to support already running IP services. The ICN network consists of nodes that follow the PURSUIT architecture~\cite{Trossen072012}\footnote{Notably, although our ICN assumptions are considerably satisfied by the PURSUIT architecture, any other architecture can adapt our solution with minor modifications, provided it can support a centralised management function and a source-routed forwarding plane.}, supporting an ICN publish/subscribe paradigm. The ICN network requires the deployment of centralized network management entities, which implement rendezvous, topology and resource management. In this work such functions are realized by a (logically) centralized control node \ie the Topology Manager (TM). On the data plane, the ICN network employs a source-routing based forwarding plane~\cite{Jokela09}. 

The required configuration parameters so that a node can be ICN functional are: (a) \emph{Node Identifier (NID)} unique in an ICN, (b) \emph{Link Identifier (LID)} unique for each point-to-point ICN forwarding connection, formed as a fixed-size Bloom filter-based vector~\cite{bloom1970space}, (c) \emph{Internal LID (iLID)} used for node-internal ICN processes, and (d) FID from an ICN node to the TM \emph{(TMFID)} following a source routing approach~\cite{Jokela09}.

\subsection{ICN Bootstrapping}

We base the ICN bootstrapping protocol to to the Dynamic Host Configuration Protocol (DHCP) principles \ie new hosts broadcast their existence to their local network and request identifiers. Centralized entities with full knowledge of the network assign and offer unique resources. ICN extends this base functionality into wider area networks. Nodes bootstrap with default node and link identifiers and send broadcast requests to their attached nodes. If a node is already part of an ICN topology, it will respond with its TMFID. Subsequently, bootstrapping nodes can directly request unique identifiers from the TM, receiving an appropriate offer if there are available resources. Finally, nodes and TM will acknowledge the assignments, and update their configuration. 

To support this functionality, the following messages and their semantics are defined: (a) \emph{DiscoveryRequest}, the first message that a new node broadcasts to its attached neighbors, (b) \emph{DiscoveryOffer}, the response sent by an attached ICN node providing its \emph{TMFID}, (c) \emph{ResourceRequest}, the request sent by the new node to the TM for a unique NID and LID, (d) \emph{ResourceOffer}, the response of the TM offering unique identifiers. It includes the triple \emph{\{NID, LID, iLID\}}, (e) \emph{OfferAccepted}, the acknowledgement of the new node to the TM accepting the offered resources, (f) \emph{ResourceAccepted}, the acknowledgement of the TM to the new node, accepting its attachment to the ICN nodes, and (g) \emph{Update}, the notification to the attached neighbor, to update its configuration with the new connection. 

 \begin{figure}[t!]
\centering
\includegraphics[width=80mm]{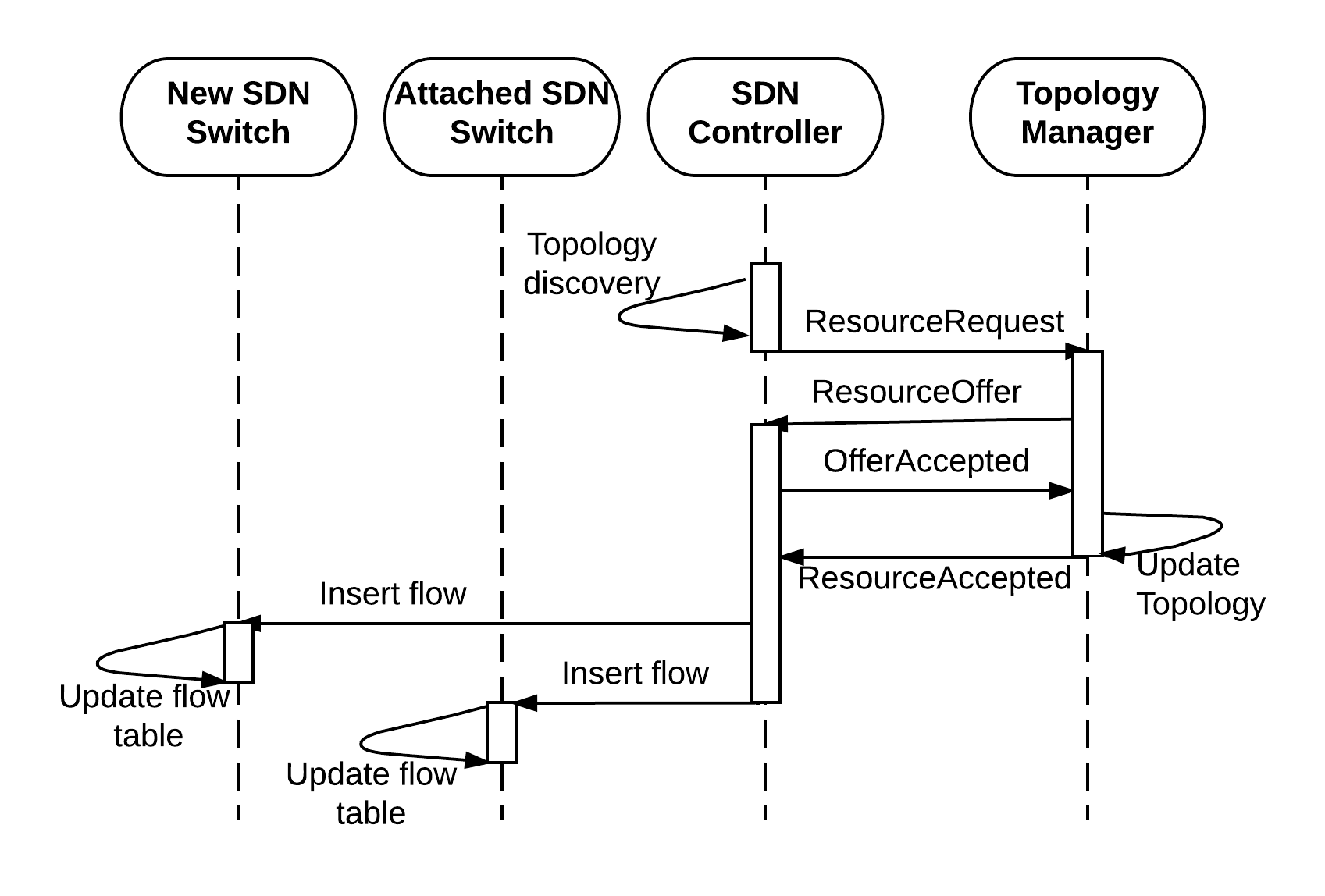}
\caption{Extending an ICN topology with a new SDN switch. \label{fig:new-switch}}
\end{figure}
\section{ICN-SDN Inter-Operability}
\vspace{-0.2pt}
ICN topologies can be realized and managed over SDN-based core networks operating legacy IP services, without modifying any existing protocol and disrupting running applications. SDN switches inherently support ICN over SDN switching using the mechanism described in~\cite{Reed15}. In this approach, the OpenFlow protocol is not modified, but utilized to support the required operations. To facilitate ICN-over-SDN topologies, an ICN-specific application is deployed on the SDN controller, directly interfacing to the ICN TM to request and configure resources. 

On the forwarding fabric level, SDN switches cannot handle any publish/subscribe operations so as to perform bootstrapping protocol functions. We overcome this constraint by configuring SDN switches to forward unknown ICN packets to the controller and be notified about the actions to perform. Consequently, the SDN controller performs some of the operations of our bootstrapping protocol, on behalf of the SDN switches, utilizing its direct interface with the TM. The TM handles those messages in the same way it handles similar bootstrapping protocol messages from ICN nodes.

\subsection{Bootstrapping an SDN Switch}
New SDN switches attach to the ICN network via previously connected SDN switches. 
When an SDN switch is attached to a ICN-enabled SDN switch, the SDN controller is notified through its internal SDN topology monitoring function as shown in Figure~\ref{fig:new-switch}. The controller acts on behalf of the switch and sends a \emph{ResourceRequest} message to the TM over the ICN-SDN interface. The TM assigns a NID and a LID (provided that resources are available) and sends a \emph{OfferAccepted} message to the Controller. A NID is not configured or stored on the switch, but is assigned for TM's topology management purposes. The Controller replies with an \emph{ResourceAccepted} message. Finally, the controller receives the \emph{ResourceAccepted} message from the TM and inserts the appropriate arbitrary bitmask flow rules to both the new and its attached SDN switch. Hence, both SDN switches are attached and ICN-enabled and can forward ICN traffic accordingly. 

\subsection{Bootstrapping an ICN node connected to an SDN Switch}
\begin{figure}[t!]
\centering
\includegraphics[width=80mm]{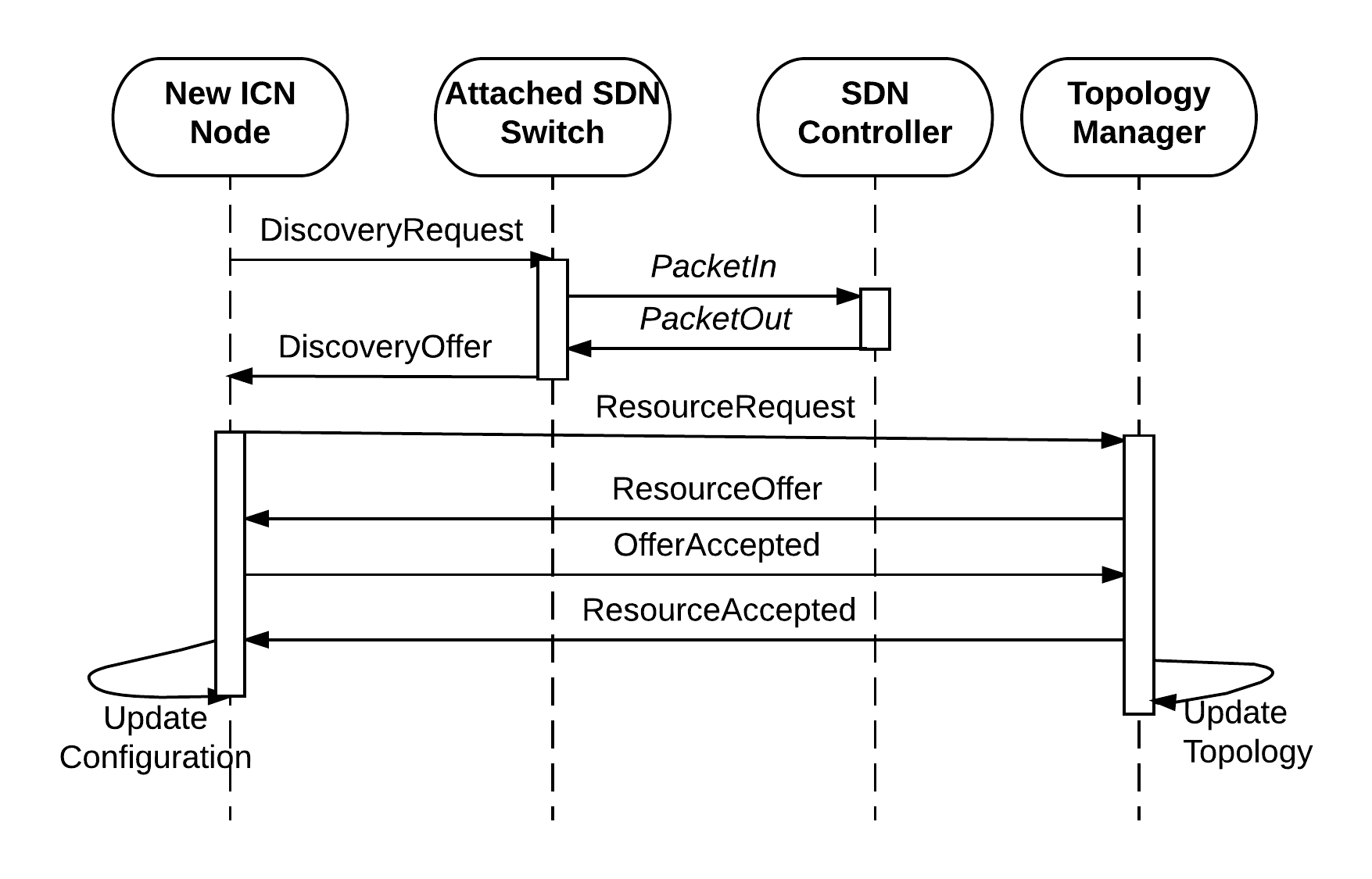}
\caption{Extending an ICN topology with a new ICN node. \label{fig:new-host}}
\end{figure}
Client and server machines, as well as proxies connected to edge SDN switches, are all expected to have typical IP connectivity. 
The bootstrapping process is executed on-demand, generating the default ICN configuration for these devices and triggering the bootstrapping mechanism illustrated in Figure~\ref{fig:new-host}. \emph{DiscoveryRequest} messages are broadcasted to their attached SDN switch, aiming to attach to the ICN topology and receive unique resources to function properly. 
The attached SDN switch receives the \emph{DiscoveryRequest} message, but does not have flow rules configured to forward it to a particular port, thus forwards it to its respective SDN controller. The message is sent to the controller encapsulated in an OpenFlow \emph{PacketIn} message. The ICN application on the controller is able to calculate the FID to reach the TM and includes it into a \emph{DiscoveryOffer} message, encapsulated in an OpenFlow \emph{PacketOut} message, which is eventually sent to the new node via the attached SDN switch. This message contains the best TMFID and the new node uses it to send a \emph{ResourceRequest} message directly to the TM. These messages traverse the already configured flow rules in SDN switches, hence no additional interaction with the SDN controller is required.
The \emph{ResourceOffer}, \emph{OfferAccepted} and \emph{ResourceAccepted} messages are exchanged between the new node and the TM. Upon successful assignment of the NID, iLID and LID to the new ICN node, the new node replaces the temporary identifiers with the unique ones obtained from the TM. The TM updates its topology graph, hence the new node becomes fully ICN functional. 

\subsection{ICN-SDN interface}
With a flexible and scalable bootstrapping mechanism in place, the monitoring of SDN switches and hosts deployed as ICN nodes becomes a necessity. The already defined ICN-SDN interface used for the bootstrapping protocol is extended to serve this purpose, as a complete management interface between the two architectures. 
The SDN controller has internal link status monitoring mechanisms, which are exploited by the deployed ICN application to be notified on the addition, update or removal of ICN nodes. The TM has to be informed of such events to enable its failover mechanisms and ensure seamless path management and packet forwarding. 
When the TM receives such a message, it identifies the path segments which are affected by this change and enables resilience mechanisms for seamless packet forwarding. The TM also supports traffic engineering functions to select the best available path in terms of link load when requested~\cite{Reed12}. Therefore, the SDN controller has to periodically send ICN link statistics to the TM to support such functionalities.

\vspace{-1pt}
\section{Evaluation}
\vspace{-1pt}
\subsection{Implementation and Setup}
To validate the feasibility of our design, a prototype has been implemented. 
Opendaylight\footnote{https://www.opendaylight.org} has been selected as the SDN controller on which to base our implementation since it is the most prominent open-source SDN platform. The required, OpenFlow standard compliant, arbitrary bitmask match support has been implemented and provided to the Opendaylight community\footnote{https://wiki.opendaylight.org/view/Simultaneous_Release/Beryllium/SR3/Release_Notes}. The ICN-aware SDN application is implemented as an Opendaylight Boron SR1 module. Our bootstrapping protocol implementation and topology management extension are based on the Blackadder software~\cite{Trossen12}, which realizes the PURSUIT architecture. 
All implemented modules and processes are fully functional and achieve the dynamic bootstrapping and extension of native ICN topologies over SDN networks and validate the designed protocol. They have become available as open-source in the latest release of the POINT software\footnote{https://github.com/point-h2020/point-2.0.0} under GPLv2 license.

To emulate SDN topologies,
custom Mininet\footnote{http://mininet.org} topologies have been generated and were attached to the Opendaylight ICN application. Their size varied in terms of number of nodes, links and new hosts placements to measure the bootstrapping time of topologies and new nodes. 
Measurements include the time to monitor topology changes, request and receive unique resources from the TM and configure the SDN switches with the relevant abitrary bitmask rules. On each new host, we measure how much time is required to be discovered, configured and eventually attached to an ICN network. 20 iterations per case were performed, and average values are presented with 99\% confidence level. 
\vspace{-1pt}
\subsection{Results}

Our mechanism does not impose any additional communication delay during content forwarding, hence the evaluation of our mechanism focuses on: (a) the bootstrapping time of new links added to the topology, and (b) the time required for a host to become ICN functional. 
Initially, the topology bootstrapping time for topologies with differing number of links is measured. The number of nodes is not important in our evaluation, since node identifiers are not required for forwarding. 
On the other hand, each link requires a unique identifier which is translated into a configured OpenFlow rule. 
Results show that topology formation time increases linearly as the number of links increase. Despite aggregating multiple resource requests in one message, this is expected because there is processing overhead in resource management in the TM. Average bootstrapping times for 10 to 60 links require from 70 up to 220 ms. 
Such bootstrapping time is considered insignificant since it will be performed before any content requests and data plane forwarding. 

We also investigate the potential impact of the node-TM network distance in terms of number of hops. ICN packet handling and TMFID calculation from the Opendaylight controller is expected to be the same from every host, hence the differentiating factor is the direct communication overhead between a host and the TM. Results show that bootstrapping time is not influenced by the hop count to reach the TM. For 1, 5, 10 and 20 hops, node bootstrapping time required 150, 144, 143 and 168ms respectively, only slightly differentiated due to the differing link delay. Therefore, clients and servers are expected to be rapidly attached to topologies, executing ICN applications with insignificant delays, and the relative placement of the TM within the IP/SDN network does not, significantly, impact performance.

\section{Conclusion}
In this paper we presented a scalable and dynamic framework for deploying and extending ICN topologies over SDN networks. 
With our approach, extension of the SDN nodes with ICN configuration is completely decoupled from the content request and forwarding phases and is performed proactively and dynamically. ICN operations and traffic forwarding are executed transparently to the SDN controller and switches without any message exchange. 
Additional management functionalities are introduced by our mechanism's design to facilitate traffic engineering and resilience functions for ICN. 
Initial results show that the delay to bootstrap and extend topologies is low, and the initialization of new nodes is independent of the number of hops from the central topology management function. All the aforementioned form a scalable and efficient ICN-SDN inter-operability framework, which opens a clear path for adoption of ICN by network providers. 

\section*{Acknowledgment}

This work was carried out within the project POINT, which has received funding from the European Union's Horizon 2020 research and innovation programme under grant agreement No 643990.

\bibliographystyle{IEEEtran}
\bibliography{icn-sdn-bootstrapping}
\end{document}